\documentclass[aps,epsfig,floats,article]{revtex4}

\usepackage[latin1]{inputenc}
\usepackage{graphics}
\usepackage[dvips]{graphicx}
\usepackage{epsfig}
\usepackage{hhline}
\usepackage{amsmath,amsfonts,amssymb}
\usepackage{xspace}
\usepackage{enumerate}
\newcommand{\etal}{et al.}

\newcommand{\dps}{\displaystyle}

\begin{document}

\title{\vspace{2cm} \Large{\textbf{Microscopic calculations of Hugoniot curves of neat TATB and of its detonation products}}}

\author{\large{\textbf{Emeric Bourasseau$^{\dagger}$\footnote{corresponding author: emeric.bourasseau@cea.fr}, Jean-Bernard Maillet$^{\dagger}$, Nicolas Desbiens$^{\dagger}$, Gabriel Stoltz$^{\ddagger}$}}}

\address{\vspace{.4cm} $^{\dagger}$CEA, DAM, DIF, F-91297 Arpajon, France\\
$^{\ddagger}$Université Paris Est, CERMICS, MICMAC Project-team,
INRIA-Ecole des Ponts ParisTech, \\
6 et 8 Av. Blaise Pascal, 77455 Marne-la-Vallée Cedex 2, France\\
}

\begin{abstract}
\textbf{Abstract.} We compute the Hugoniot curves of both neat TATB and its detonation products mixture using atomistic simulation tools. To compute the Hugoniot states, we adapted our "Sampling Constraints in Average" (SCA) method (Maillet \etal, \textit{Applied Math. Research eXpress} \textbf{2008}, 2009) to Monte-Carlo simulations. For neat TATB, we show that the potential proposed by Rai (Rai \etal, \textit{J. Chem. Phys.} \textbf{129}, 2008) is not accurate enough to predict the Hugoniot curve and requires some optimization of its parameters. Concerning detonation products, thermodynamic properties at chemical equilibrium are computed using a specific RxMC method (Bourasseau \etal, \textit{Phys. Chem. Chem. Phys.} \textbf{13}, 2011) taking into account the presence of carbon clusters in the fluid mixture. We show that this explicit description of the solid phase immersed in the fluid phase modifies the chemical equilibrium.
\end{abstract}

\maketitle

\section{Introduction}

To understand and model detonation phenomena, it is very important to
describe accurately the thermodynamic properties of the neat explosive
as well as its detonation product mixture.  When a material is hit by
a shock wave, its thermodynamic state changes towards a state with
higher temperature, higher density and higher pressure. The set of
admissible states which can be attained from given initial
conditions, for shocks of various strengths, is called the Hugoniot
curve (see equation~\eqref{eq:RankineHugoniot} for a precise
definition). For explosives, exothermic chemical decompositions are
triggered by the passage of a shock wave. However, the state of the
material after the propagation of the shock is a particular point of
the Hugoniot curve of the \emph{unreacted} material, the so-called ZND
state, which corresponds to a high pressure and a relatively high
temperature \cite{zeldovich40, neumann42, doring43}. From this state, exothermic chemical
reactions produce small stable molecules (such as H$_2$O, CO$_2$,
N$_2$) called detonation products. As for every thermally activated
processes, the initial thermodynamic conditions of the system before
it reacts are one of the parameters controlling the reaction
rate. Therefore, the determination of the temperature of the material after the 
passage of the shock wave,
and more generally of the complete equation of state (EOS) of the
neat explosive, is a key point for predicting the first step
of the molecular decomposition. Moreover, the energy released by a given
explosive can be estimated from the knowledge of the isentropic curve
of the detonation products mixture in the pressure-volume plane. Thus, important efforts have also been dedicated in the past decades to
measure and predict the equilibrium properties of the detonation products
in the high pressure and high temperature regime. Unfortunately,
measurements of thermodynamic properties of both shocked material and
detonation product mixtures are relatively scarce. As a consequence,
following a similar work on liquid nitromethane and other
energetic materials~\cite{bourasseau07,desbiens09}, we compute here
the Hugoniot curves of TATB and of its detonation products
using techniques and models from statistical physics.

To this end, we adapted the "Sampling
Constraints in Average" method (SCA) proposed in~\cite{maillet09} to our molecular Monte
Carlo simulation tool. This method aims at sampling microscopic configurations
of a system, consistent with a thermodynamic ensemble (canonical, isobaric-isothermal,
grand-canonical, etc), and such that some constraints are satisfied in average. 
A typical application is the determination of the temperature of a system 
given its average energy and its density.
Here, we use SCA to compute the Hugoniot curves of both the neat
explosive and the detonation product mixture. The composition at
chemical equilibrium of the detonation product mixture, is obtained
by the RxMC method~\cite{bourasseau11}. 
It is important to take into account the presence of solid carbon clusters
in the detonation product mixture. The RxMC method models
carbon clusters as mesoparticles embedded in the mixture and therefore 
allows us to calculate thermodynamic properties of a heterogeneous liquid/solid system at
chemical equilibrium. 

This paper is organized as follows. In Section~\ref{sec:SCA}, we briefly review 
the method to sample constraints in average, in the particular
case of the computation of Hugoniot curves. We then describe 
the system we consider and the sampling method used to sample its microscopic configurations
in Section~\ref{sec:description}. Numerical results are then 
presented in Section~\ref{resultats}. Our findings are finally summarized 
in the conclusion (Section~\ref{sec:conclusion}).

\section{Sampling Constraints in Average}
\label{sec:SCA}

\subsection{Description of the method}

We present here the SCA method applied to the computation of Hugoniot
curves when a Monte Carlo sampling method is used (see~\cite{maillet09} for a
more general presentation and other possible applications).  

\subsubsection{Hugoniot curves}
The
Hugoniot curve is the ensemble of accessible thermodynamic states that
a system can reach from a given initial state after the passage of a
shock wave. The thermodynamic quantities of a material in the initial
unshocked state and the final shocked state are related by the
Rankine-Hugoniot relation:
\begin{equation}
  \label{eq:RankineHugoniot}
  E - E_0 = \frac{1}{2}(P+P_0)(V_0-V),
\end{equation}
where $P$, $V$, and $E$ are respectively the pressure, volume and
energy of the system in the shocked state (at temperature $T$), and
$P_0$, $V_0$, and $E_0$ are the pressure, volume and energy of the
system in the initial state (pole).

Macroscopic relations such as~\eqref{eq:RankineHugoniot} can be
obtained by averaging functions of the microscopic state of the
system, as predicted by the laws of statistical physics.  In this
study, we restrict ourselves to the NPT ensemble, but generalizations
to other thermodynamic ensembles are straightforward.  We denote by
$(q,V)$ the configuration of the system: $q=(q_1,\dots,q_A)$ are the
positions of the $A$~particles, and $V$ is the volume of the simulation
box.  The potential energy of the system is denoted by $E_{\rm
  pot}(q)$.  A reformulation of~\eqref{eq:RankineHugoniot} consists in
replacing $E$ by the average microscopic energy obtained by
averaging~$E_{\rm pot}$ in the thermodynamic ensemble at hand and
adding the kinetic contribution, and replacing $V$ by the average
volume at the given temperature and pressure.  Therefore,
\eqref{eq:RankineHugoniot} should be understood as
\begin{equation}
\label{eq:RankineHugoniot_phys_stat}
\langle H \rangle_{P,T} - \langle H \rangle_{P_0,T_0} =
\frac{1}{2}(P+P_0)\Big( \langle V \rangle_{P_0,T_0} -\langle V
\rangle_{P,T} \Big),
\end{equation}
where $\langle \cdot \rangle_{P,T}$ denotes an isobaric-isothermal average
at fixed pressure~$P$ and temperature~$T$.

\subsubsection{Previous methods to compute Hugoniot curves with Monte Carlo sampling method}
\label{AEEOS}
Hugoniot curves were previously computed with the AE-EOS method
\cite{erpenbeck92, bourasseau07, hervouet08}. In this method, the
pressure~$P$ of the state on the Hugoniot curve is fixed, and the
temperature such that~\eqref{eq:RankineHugoniot_phys_stat} is
satisfied is determined iteratively, using Newton's algorithm. More
precisely, denote by $T^n$ the approximation of the temperature~$T$ at
the $n^{th}$ step. A short NPT simulation at temperature~$T^n$ and
pressure~$P$ is run, and an approximation to the average
\begin{equation}
H_{\rm g}(T^n) = \langle H \rangle_{P,T^n} -
E_0 - \frac{1}{2}(P+P_0)\Big(V_0-\langle V \rangle_{P,T^n}\Big)
\label{Hg}
\end{equation}
is computed. The derivative of $H_{\rm g}$ with respect to~$T$ is
then estimated at the temperature~$T^n$ using a finite difference, and
a new temperature is obtained as
\[
T^{n+1} = T^n - H_{\rm g}(T^n) \frac{T^n - T^{n-1}} {H_{\rm g}(T^n)-H_{\rm g}(T^{n-1})}.
\]
To initialize the algorithm, $T^0$ is chosen arbitrarily, and $T^1$ is obtained by adding a given $\Delta T$ to $T^1$. The temperature $T^n$ converges to the
temperature $T_{H_{\rm g}}$, which corresponds to $H_{\rm g}(T_{H_{\rm
    g}}) = 0$. The AE-EOS method gives good results, and allowed us to
obtain the Hugoniot curves of various energetic materials and their
detonation products mixtures. 
However, its major drawback is that
the modifications of the temperature do not easily allow to perform a statistical
ensemble average on the full calculation. 
At the end of the AE-EOS calculation, a usual NVT or NPT simulation at temperature $T = T_{H_{\rm g}}$
is needed to obtain the direct and derivative thermodynamic properties.
Therefore, a lot of computational time is wasted in the process.

\subsubsection{Sampling constraints in average}

To get rid of this important drawback of the AE-EOS method, we have recently
developed a more efficient method (called SCA) which allows to sample 
microscopic configurations with a constraint satisfied in average
during the simulation. This method was developed in a molecular dynamics
framework~\cite{maillet09}. We extend it here to Monte Carlo simulations. 
Applied to the Hugoniot calculation case,
it allows to determine the temperature $T_{H_{\rm g}}$ such that
$H_{\rm g}$ is indeed equal to zero in average, and it moreover allows to
correctly explore the phase-space configurations consistent with the appropriate
thermodynamic ensemble at this temperature.
Thus, a single simulation gives the Hugoniot temperature
and the thermodynamic properties of the system at this point. The convergence
of this method was studied from a mathematical viewpoint in~\cite{maillet09}.

Let us now present a brief description of the adaptation of SCA to a Monte
Carlo sampling method such as standard Metropolis-Hastings algorithms~\cite{MRRTT53,Hastings70}.
Heuristically, the idea is to select a control 
variable (the temperature $T$ here), to decompose its values
into bins, and to accumulate 
the instantaneous values of the property under investigation 
($H_{\rm g}$ in our application) every time
the control variable falls in the corresponding bin.
This allows to construct estimates of the average property as a function of 
the control variable. To this end, 
a standard NPT simulation is performed, with an additional dynamical
update of the temperature depending on the observed values of the average
property which should be constrained.
The key point is to determine how the temperature is
changed. 
In our application, $H_{\rm g}$ is positive when $T > T_{H_{\rm g}}$ and 
negative when $T < T_{H_{\rm g}}$. As our goal is to find the temperature $T_{H_{\rm g}}$
for which $H_{\rm g}$ is zero, it is natural to resort to the following temperature update:
\[
T^{n+1} = T^n - \alpha \langle H_{\rm g} \rangle_{P,T^n}.
\]
In fact, the average value $\langle H_{\rm g} \rangle_{P,T^n}$ is not known exactly,
and should be replaced with its approximation obtained from the histogram
for the bin corresponding to $T^n$. It corresponds to an average computed along the whole 
trajectory, which is therefore more and more accurate as the
simulation goes on. 

\begin{figure}[!h]
\begin{center}
\epsfig{file=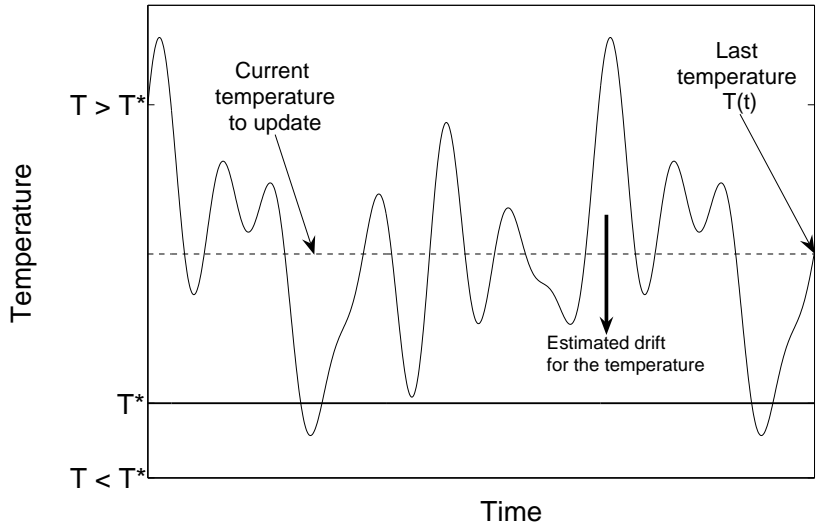,width=12cm, angle=0}
\end{center}
\caption{Schematic representation of the SCA algorithm.} \label{fig:cartoon}
\end{figure}

More precisely, consider 
a temperature grid $T^i = T_{\rm min} + i \Delta T$ with $i$ taken between $0$ and $M$, and $T^M = T_{\rm max}$.
Denote by $I(T)$ the function which returns the 
index of the bin corresponding to the temperature~$T$, and by 
MC$_{P,T}(q,V)$ the Monte Carlo algorithm which returns a new configuration $q',V'$
from a previous one, for a given pressure $P$ and a given temperature~$T$.
The SCA algorithm then schematically reads as follows:
\[
\left\{ \begin{array}{l}
(q^{n+1},V^{n+1}) = \mathrm{MC}_{P,T^n}(q^n,V^n), \\
\dps T^{n+1} = T^n - \alpha \frac{\dps \sum_{m = 0}^{n+1} \left( 
E(q^m,V^m) - E_0 + \frac12 (P+P_0)(V_0-V^m) \right)
\delta_{I(T^m)=I(T^n)}}
{\dps \sum_{m = 0}^{n+1} \delta_{I(T^m)=I(T^n)}}.
\end{array} \right.
\]
See also Figure~\ref{fig:cartoon} for a cartoon representation of the procedure.
In fact, for Monte Carlo sampling methods, it is convenient to repeat several times (say, $N$ times)
the update of the configuration before updating the temperature.

It is expected that $T^n$ converges to $T_{H_{\rm g}}$. As for the continuous dynamics
considered in~\cite{maillet09}, such a convergence can be expected only if the parameters of the algorithm are well chosen -- although it is very easy to find 
satisfying values of these parameters, see Section~\ref{sec:application_modele}.
The two parameters that control the efficiency of the algorithm are $\alpha$ and $N$:
\begin{enumerate}[(i)]
\item The quantity $\alpha$ controls the rate at which the temperature is updated.
If $\alpha$ is too small, the temperature changes very slowly, leading to an inefficient
algorithm. If $\alpha$ is too large, the variations are too brutal and the system
is driven out of equilibrium. This may lead to numerical instabilities.
A practical way of choosing $\alpha$ is therefore to start from a 
conservatively small value, and increasing it until some numerical instabilities
are observed. This can be done using very short preliminary computations.
Besides, our experience is that the precise value of $\alpha$ is not very important,
and the range of admissible $\alpha$ is robust with respect to the changes
in the thermodynamic conditions.
In any case, the value given to
$\alpha$ does not change the final result, it only has an
influence on the efficiency of the convergence.
\item The number of steps $N$ between two temperature updates. Actually, in the original
Molecular Dynamics approach~\cite{maillet09}, the temperature is changed at every time step. However, in a Monte
Carlo simulation, the system may only be slightly modified at each Monte Carlo step,
and it appears more convenient to let the system equilibrate a little while
at a given temperature before changing it again. As a consequence, a reasonable value
for $N$ could be an entire Monte Carlo cycle between two temperature
changes.
\end{enumerate}

\subsection{Application to a model system}
\label{sec:application_modele}

To understand the influence of the parameters~$N$ and~$\alpha$, we performed several simulations with this method on a test system composed of 400 point-like particles interacting with a standard pairwise Lennard-Jones 6-12 interaction potential depending only on the relative distances between the particles. A cut-off equal to the half of the box length has been used to reduce the computing time. Periodic boundary conditions together with long range corrections have been used.
%\[
%v(r_{ij}) = 4 \epsilon \left( \left( \frac{\sigma}{r_{ij}} \right)^{12} -  \left( %\frac{\sigma}{r_{ij}} \right)^{6} \right)
%\]

Parameters have been taken arbitrarily equal to $\epsilon = 120.0$ K and $\sigma = 3.40~\AA$. 
We computed the temperature for which $H_{\rm g} = 0$, at $P = 1$~GPa, with initial thermodynamic conditions arbitrarily chosen as: $E_0$ = 345 J.g$^{-1}$, $V_0$ = 5.848 $10^{-7}$~m$^3$.g$^{-1}$ and $P_0$ = 1 $10^5$ Pa. Monte Carlo moves in these test simulations are translation moves (95 \%) and isotropic volume changes (5 \%). Translation distances and volume changes coefficients are selected randomly under maximum values that are updated every 10000 MC moves in order to obtain 40 \% of accepted moves. A standard Metropolis scheme is used to accept or reject new configurations. Finally, around 1000 MC moves are needed to complete a Monte Carlo cycle.\\

\begin{table}[h!]
\centering
\begin{tabular}{|c|c|c|c|c|c||c|c|}
\hline
 & \multicolumn{5}{c||}{Sampling Constraints in Average (New method)} & \multicolumn{2}{c|}{AE-EOS (Old method)} \\
\hline
\hline
~~~~~~N~~~~~~ & ~~~~1000~~~~ & ~~~~1000~~~~ & ~~~~1000~~~~ & ~~~~100~~~~~ & ~~~10000~~~~ & ~~~10000~~~~ & ~~~100000~~~ \\
$\alpha$ (K.g.J$^{-1}$)& 1. & 0.1 & 0.01 & 0.1 & 0.1 & - & - \\
\hline
\hline
$T_{H_{\rm g}}$ & 1285.7 & 1284.0 & 1284.9 & 1284.7 & 1284.8 & 1279.2 & 1285.6 \\
(K) & $\pm$ 0.4 & $\pm$ 0.04 & $\pm$ 0.06 & $\pm$ 0.2 & $\pm$ 0.2 & $\pm$ 0.0 & $\pm$ 1.8 \\
\hline
$H_{\rm g}$ & -0.06 & 0.13 & 0.14 & -0.38 & 0.06 & -2.22 & 1.40 \\
(J.g$^{-1}$) & $\pm$ 0.53 & $\pm$ 0.42 & $\pm$ 0.57 & $\pm$ 0.56 & $\pm$ 0.61 & $\pm$ 0.46 & $\pm$ 1.53 \\
\hline
\end{tabular} \caption{Comparison between results obtained on the test system with the SCA method (New method), and the AE-EOS method.}
\label{ConvTab}
\end{table}

Table~\ref{ConvTab} shows the results obtained for different values of $\alpha$ and $N$. We also show in this table the results obtained with the AE-EOS method. In the two cases, the parameter N corresponds to the number of MC moves performed between two temperature changes. A total of 10$^7$ MC moves have been performed in all cases, and statistical errors have been estimated using blocks averages over the 5.10$^6$ last iterations. Concerning the SCA method, it appears that the obtained temperature does not depend on the parameters, as expected. The "Hugoniot" temperature is equal to 1285 $\pm$ 1 K. Moreover, the constraint $H_{\rm g} = 0$ is respected in average for all the parameter sets considered. Note that in such a system, the total energy is about several tens of J.g$^{-1}$, so that the statistical uncertainties observed with both methods are very small. However, the results from the new method are more accurate and reliable that those obtained with the AE-EOS method at a fixed computational cost. Moreover, it appears that AE-EOS method can converge to wrong values of the temperature: in the example corresponding to N = 10000, the temperature seems extremely well converged with a very reduced statistical error, but with a corresponding $H_{\rm g}$ value not equal to zero.\\

\begin{figure}[!h]
\begin{center}
\epsfig{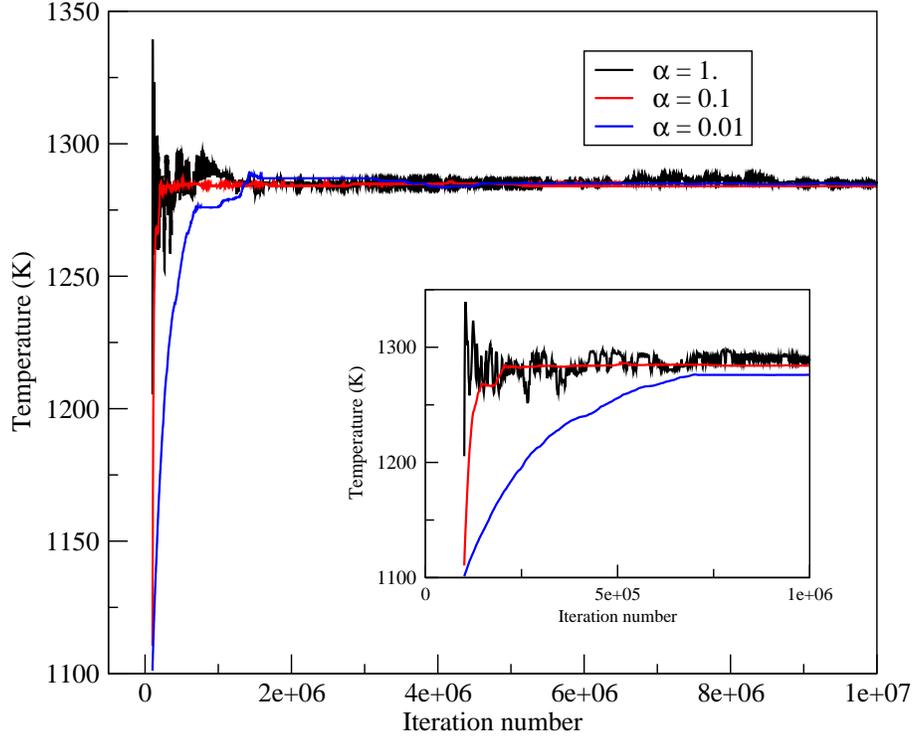}
\end{center}
\caption{Evolution of the temperature during the simulation for various values of the parameter~$\alpha$ (in K.g.J$^{-1}$). The value~$N$ is kept constant and equal to 1000. The insert corresponds to a zoom on the first $10^6$~iterations.} 
\label{ConvVsa}
\end{figure}

Figure \ref{ConvVsa} shows the evolution of the temperature observed during the simulation, for various values of $\alpha$, when $N$ is kept constant and equal to~1000.
As shown in Table~\ref{ConvTab}, the temperature obtained at the end of the simulation does not depend on the parameter~$\alpha$. However, too high a value of $\alpha$ triggers oscillations of the temperature around the target temperature, leading to higher statistical uncertainties. 
On the opposite, too small a value of $\alpha$ leads to a very slow and inefficient convergence. 

\begin{figure}[!h]
\begin{center}
\epsfig{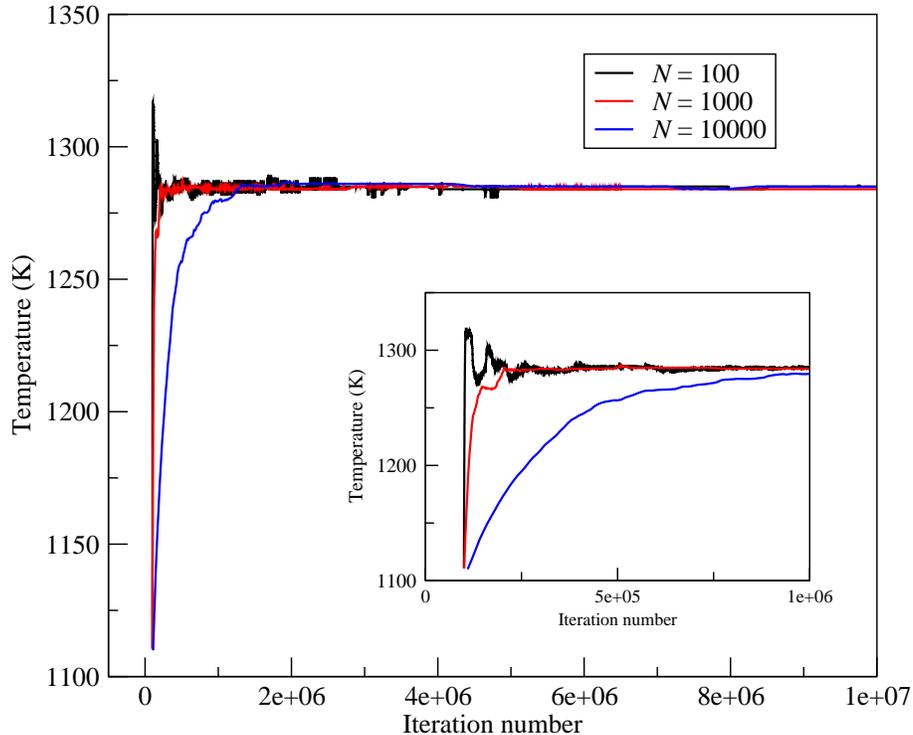}
\end{center}
\caption{Evolution of the temperature during the simulation for various values of the parameter~$N$. The value~$\alpha$ is kept constant and equal to 0.1 K.g.J$^{-1}$. The insert corresponds to a zoom on the first $10^6$~iterations.} 
\label{ConvVsN}
\end{figure}

Figure \ref{ConvVsN} shows the evolution of the temperature observed during the simulation, for various values of~$N$, keeping $\alpha$ constant and equal to 0.1~K.g.J$^{-1}$.
Once again, the limiting value of the temperature does not depend on the value of~$N$. However, it seems that too high a frequency of temperature updates (corresponding to a small value of $N$) can lead to instabilities, whereas too high a value of $N$ slows down the convergence.

%--------------- description of the system -----------------
\section{Description of the system}
\label{sec:description}

We present successively the Monte Carlo method used to sample the configurations 
of the system (Section~\ref{sec:RxMC}), and the atomistic description of the interactions
(Section~\ref{sec:force_fields}).

\subsection{RxMC with mesoparticle}
\label{sec:RxMC}

The goal of the RxMC method is to predict equilibrium properties of multi-component systems constrained by chemical equations. Microscopic configurations consistent with the chemical equilibrium can be sampled by resorting to a specific thermodynamic ensemble: the Reaction Ensemble.
RxMC is a well-established molecular-level simulation method to sample the Reaction Ensemble and it has found applications in a large variety of problems. Interested readers are referred to previously published articles on the method and its application. In particular, Turner {\it et al.} have published a review of the method and its applications prior to 2008 \cite{turner08}, and some references for more recent applications are~\cite{bourasseau07, lisal08, lisal09, bourasseau11}. The complete definition of this ensemble, including an expression of its probability density, was given first by Smith and Triska \cite{smith94}. J.K.
Johnson has also given a very nice derivation of the Reaction Ensemble acceptance
probability~\cite{johnson94}. Another interesting early reference is the pioneering work of M.S. Shaw, who proposed a method similar in nature to RxMC to simulate chemical equilibrium of molecular mixtures~\cite{shaw91,shaw01}. 

The RxMC method is based on a particular Monte Carlo move, the reaction move. For a given chemical equilibrium, this move consists in changing the chemical composition of the system by deleting reactant molecules, and inserting products molecules in order to keep constant the number of atoms of each species. Of course, the move has to be performed in forward and backward directions to ensure micro-reversibility. The averaged chemical composition of the simulation corresponds to the equilibrium composition of the system.

In the reference \cite{bourasseau11}, we give a detailed description of the original method, and we also describe the new method we have proposed to perform RxMC simulation of detonation product mixtures that include solid carbon clusters in the fluid phase. 
It is indeed very important to explicitly model this phase of carbon in order to obtain 
correct simulation results
since the chemical equilibrium of a single heterogeneous system is thermodynamically different from considering two homogeneous separated systems ({\it i.e.} liquid and solid) in equilibrium. Significant differences have been found between thermochemical calculations (considering separated phases) and explicit microscopic simulations of the heterogeneous system \cite{bourasseau11}.

The method described in \cite{bourasseau11} models the solid phase as a mesoparticle representative of a cluster of N$_{\rm C}$ carbon atoms (denoted by MP$_ {N_{\rm C}}$ in the sequel), immersed in the reacting fluid. In this case, the mesoparticle can be considered as a single rigid molecule, and the chemical equations we consider to take into account the chemical equilibrium between the fluid and the solid phases are similar to the following example:
\begin{equation}
2 \, \mathrm{CO} + \mathrm{MP}_{N_{\rm C}} \rightleftarrows \mathrm{CO}_2 + \mathrm{MP}_{N_{\rm C}+1}
\end{equation}
In the forward direction, the corresponding reaction move consists in deleting two CO molecules, and the mesoparticle of N$_{\rm C}$ carbon atoms, and inserting a CO$_2$ molecule and the new mesoparticle MP$_{N_{\rm C}+1}$. Of course, an efficient way to proceed is to replace the old mesoparticle by the new mesoparticle, which corresponds in practice to changing the mesoparticle volume according to the addition of a carbon atom. This move is no longer completely random, and the bias has to be corrected. This is easily done by performing deletions, insertions and mesoparticle replacements always in the same order (see~\cite{bourasseau07} for further precisions).

It is possible to write the expression of the acceptance probability of such a move from the standard equations of the RxMC ensemble (see~\cite{bourasseau11}, which is dedicated to the precise description of this method). Denoting by $s$ the number of chemical species (where the $s$th species is the solid carbon), it holds
\begin{equation}
\label{paccRxMCMP}
\begin{split}
P_{\rm acc} = \min \Biggl( 1, &\left( P_0 \beta V \right)^{\xi \bar{\nu}} \cdot \exp \left( -\xi \frac{ \sum _{i=1}^{s-1}  \nu_i \Delta_{\rm f} G^0_i (T)}{RT} \right) \cdot \\
& \exp \left( -\xi \frac{\nu_{{\rm C}_{\rm Sol}} \Delta_{\rm f} G_{{\rm C}_{{\rm Sol}}} (T,P)}{RT} \right) \cdot  \prod_{i=1}^s \frac{N_i!}{\left(N_i + \xi \nu_i \right)!} \exp
\left( -\beta \Delta U \right) \Biggr),
\end{split}
\end{equation}
where $P_0$ is the standard pressure, $\beta = 1/(k_{\rm B}T)$, V is the volume of the system, $\xi$ is the move direction ($\xi = +1$ if the reaction
move is performed in the forward direction, or -1 if it is performed backward), $\bar{\nu} = \sum _{i=1}^s \nu_i$ with $\nu_i$ the stoichiometric coefficient of the species $i$ involved in the reaction (for the solid carbon, the notation $\nu_s = \nu_{{\rm C}_{\rm Sol}}$ is used), $\Delta_{\rm f} G^0_i (T)$ is the standard Gibbs free energy of formation of the $i$th species at temperature~$T$, $\Delta_{\rm f} G_{{\rm C}_{\rm Sol}} (T,P)$ is the Gibbs free energy of formation of a mole of solid carbon cluster at temperature~$T$ and pressure~$P$, $N_i$ is the number of molecules of the $i$th species in the system, and $\Delta U$ is the energy difference between the old and new configurations.

It is interesting to note that the only input data needed in the reaction move are the free energies of formation $\Delta _{\rm f} G_i^0 (T)$ of the different species (for $ i\leq s-1$) and $\Delta_{\rm f} G_{C_{\rm Sol}} (T,P)$. All these quantities can be obtained from experimental databases or numerical studies \cite{bourasseau11}. It is also possible to compute a chemical equilibrium involving several chemical reactions. To this end, we add a preliminary step consisting of randomly choosing the chemical reaction before each reaction move. Finally, using in addition the usual volume change of standard MC simulations, it is also possible to simulate a chemical equilibrium at constant pressure.

Combining translation, rotation and reaction moves, and volumes changes, it is possible to simulate a chemical equilibrium at a given temperature and a given pressure. Statistical biases can be resorted to in order to improve the acceptance probability of insertion in dense phases. We employed to the pre-insertion bias method~\cite{bourasseau02}, already implemented in our MC code. An alternative (and similar) technique is the cavity bias sampling method of~\cite{brennan05}. The biases applied to the reaction move consist in inserting the first product molecules at the empty locations of the previously deleted reactant molecules. If other product molecules have to be inserted ({\it i.e.} if $\bar{\nu} > 0$), the insertion is performed at a pre-selected location. The details of the algorithm and the expression of the acceptance probability when using the pre-insertion bias are presented in \cite{bourasseau07}.

\subsection{Force fields}
\label{sec:force_fields}

To perform all atom simulations of neat TATB, two force fields are available in the literature, one from Gee {\it et al.}, proposed in 2004 \cite{gee04}, and another from Rai {\it et al.}, proposed in 2008 \cite{rai08}. The first force field is not suited for Monte Carlo simulations, since molecules are supposed to be flexible, which imposes some additional expensive internal relaxation moves. As intramolecular hydrogen bonds and benzene cycles impose a relatively high rigidity of the molecular structure, we believe that the rigid molecule approximation is valid in this case. Moreover, atomic charges given by~\cite{gee04} seem to be unreliable. The second force field describes rigid molecules and contains more realistic atomic charges. We therefore decided to use the potential given in~\cite{rai08} in our simulations.
This potential has been developed from the TraPPE force field of aniline and nitrobenzene molecules, whose molecular structures are close to the TATB structure. Parameters of intermolecular atomic Lennard-Jones 6-12 potentials have been transferred directly from those two molecules whereas ab initio calculations have been used to set the molecular geometry and the atomic charges. Table \ref{ParamTATB} gives the complete list of parameters used to model the neat TATB molecule.

\begin{table}[h!]
\centering
\begin{tabular}{|c|c|c|c|}
\hline
 & ~~$\sigma$ ($\AA$)~~ & ~~$\epsilon$ (K)~~ & ~~q (e)~~ \\
\hline
~~C (-NO$_2$)~~ & 3.60 & 30.7 & -0.242 \\
C (-NH$_2$) & 3.60 & 30.7 & +0.408 \\
N (-O$_2$) & 2.90 & 30.0 & +0.007 \\
N (-H$_2$) & 3.25 & 160.0 & -0.738 \\
O & 2.70 & 42.0 & -0.104 \\
H & 0.50 & 12.0 & +0.386 \\
\hline
\end{tabular} \caption{Lennard-Jones 6-12 parameters and atomic partial charges used to model the atomic force centres of neat TATB.}
\label{ParamTATB}
\end{table}

The experimental molecular structure of TATB has been measured at ambient temperature by Cady and Larson in 1965 \cite{cady65}. The experimental geometry is given in Table~\ref{geoexp}. Dihedral angles are equal to~0 except $\widehat{CCNO}$ which is equal to 12$^{\circ}$. Experimentally, the TATB molecule is neither absolutely plane, nor absolutely symmetric. The parameters used in our simulations are given in Table~\ref{geotheo}. All dihedral angles are set to~0.

\begin{table}[h!]
\centering
\begin{tabular}{cc|cc}
\multicolumn{2}{c|}{$d_{X-Y}$ ($\AA$)} & \multicolumn{2}{c}{$\widehat{X-Y-Z}$ ($^{\circ}$)} \\
\hline
 & & ~~~~C-C(-NH$_2$)-C~~~~ & 117.9 \\
 C-C & ~~1.442~~ & C-C(-NO$_2$)-C & 122.0 \\
 ~~~~C-N(-H$_2$)~~~~ & 1.314 & C-N-O & 121.0 \\
 C-N(-O$_2$) & 1.419 & O-N-O & 117.9 \\
 N-O & 1.243 & C-C-N(-O$_2$) & 119.0 \\
 & & C-C-N(-H$_2$) & 121.1 \\
\end{tabular} \caption{Experimental molecular structure of TATB at ambient temperature \cite{cady65}.}
\label{geoexp}
\end{table}

\begin{table}[h!]
\centering
\begin{tabular}{cc|cc}
\multicolumn{2}{c|}{$d_{X-Y}$ ($\AA$)} & \multicolumn{2}{c}{$\widehat{X-Y-Z}$ ($^{\circ}$)} \\
\hline
 & & ~~~~C-C(-NH$_2$)-C~~~~ & 118.91 \\
 C-C & ~~1.437~~ & C-C(-NO$_2$)-C & 121.09 \\
 ~~~~C-N(-H$_2$)~~~~ & 1.317 & C-N-O & 120.66 \\
 C-N(-O$_2$) & 1.422 & O-N-O & 118.68 \\
 N-O & 1.235 & C-C-N(-O$_2$) & 119.45 \\
 N-H & 1.014 & C-C-N(-H$_2$) & 120.54 \\
 & & H-N-C & 116.53 \\
 & & H-N-H & 126.94 \\
\end{tabular} \caption{Molecular structure of TATB used in our simulations \cite{rai08}.}
\label{geotheo}
\end{table}

We consider 8 different chemical species to represent the detonation products mixture of TATB (solid carbon, CO$_2$, H$_2$O, CO, N$_2$, H$_2$, NH$_3$ and CH$_4$). Molecules in the fluid phase have been modelled through the exp-6 potential from Fried {\it et al.} \cite{fried02}. The solid carbon has been modelled through a mesoparticle representing a carbon cluster, using the model we developed recently \cite{bourasseau11}. In this model, the inner properties of the cluster are given by an equation of state. Radius of the mesoparticle, and interaction potential between the mesoparticle and a fluid particle have been obtained from molecular dynamic simulations of all atoms carbon clusters using the LCBOPII potential \cite{chevrot09}. Details are given in reference \cite{bourasseau11}. We also show in this reference that this model is particularly well suited to represent the effect of a real carbon cluster immersed in a fluid mixture. 

%----------- results -------------
\section{Numerical results}
\label{resultats}

The preliminary study presented in Section~\ref{sec:application_modele} gives an order of magnitude for the parameters~$\alpha$ and~$N$. We performed a few small SCA preliminary runs for the TATB model at hand, and decided to use the following parameters to compute the Hugoniot curves: for neat TATB, $N$ = 2000 and $\alpha$ = 1.~K.g.J$^{-1}$, while for the detonation product mixture of pure TATB, $N$ = 20,000 and $\alpha$ = 0.2~K.g.J$^{-1}$.

\subsection{Hugoniot curve of neat TATB}

\begin{table}[h!]
\centering
\begin{tabular}{|c|c|c|c|c|}
\hline
& & Experimental & Rai~\cite{rai08} {\it et al.} & This work \\
& & Results \cite{cady65} & \cite{rai08} & \\
\hline
& $\rho _0$ (g.cm$^{-3}$ & 1.938 & 1.929 & 1.932 \\
Direct & P$_0$ (Pa) & 1.10$^5$ & 1.10$^5$ & 1.10$^5$ \\
Properties & V$_0$ (m$^3$.g$^{-1}$) & 5.16.10$^{-7}$ & 5.184.10$^{-7}$ & 5.175.10$^{-7}$ \\
 & E$_0$ (J.g$^{-1}$) & - & - & 1346.7 \\
 \hline
 & a  ($\AA$) & 9.010 & 9.05 & 9.037 \\
 & b  ($\AA$) & 9.028 & 9.04 & 9.018 \\
 cell & c  ($\AA$) & 6.812 & 6.80 & 6.802 \\
 parameters & $\alpha$ ($^{\circ})$ & 108.59 & 110.0 & 110.087 \\
 & $\beta$ ($^{\circ})$ & 91.82 & 88.2 & 88.086 \\
 & $\gamma$ ($^{\circ})$ & 119.97 & 120.2 & 120.185 \\
\hline
 & C$_P$ (J.K$^{-1}$.g$^{-1}$) & 1.00 \cite{olinger76} & - & 0.665 $\pm$ 0.01 \\ 
 & K$_0$ (GPa) & 17.3 \cite{olinger76} & - & 9.907 $\pm$ 0.29 \\ 
 Derivative & & 18.9 \cite{stevens08} & & \\
 Properties & $\Gamma$ & 0.20 \cite{olinger76} & - & 1.357 $\pm$ 0.12 \\
 & C$_s$ (m.s$^{-1}$) & 1460 \cite{cady65} & - & 2338.8 $\pm$ 44.6 \\ 
 & $\alpha_V$ ($\mu$m.m$^{-1}$.K$^{-1}$) & 304 \cite{olinger76} & - & 163.03 $\pm$ 7.6 \\ 
 \hline
\end{tabular} \caption{Direct and derivative thermodynamic properties and cell parameters of TATB calculated at ambient temperature (pole properties), and compared to Rai's results~\cite{rai08} and various experiments. C$_P$: calorific capacity at constant pressure, K$_0$: bulk modulus, $\Gamma$: Gruneisen coefficient, C$_S$: sound velocity, $\alpha_V$: thermal expansion coefficient.}
\label{ResPole}
\end{table}

We first computed the thermodynamic properties of neat TATB at the pole, 
which corresponds to $T_ 0 = 300$~K and $P_0 = 10^{5}$~Pa. The 
results are given in Table~\ref{ResPole}.
They show that the computed direct thermodynamic properties
and cell parameters are close to both experimental
results and Rai's calculations. Differences between our results and
Rai's results mainly arise from the way the
long range electrostatic interactions are taken into account: we used the reaction field
method \cite{hansen} whereas \cite{rai08} used the Ewald summation method. 
On the other hand, there are important differences in the derivative
properties between the computed values and the experimental results. 
Although the reliability of experimental measurements is not completely
guaranteed, the magnitude of the differences underlines a weakness of the
potential, and shows that the potential does not reproduce
correctly the evolution of the pressure as a function of the volume (K$_0$),
of the energy as a function of the temperature (C$_P$), and of the volume as a
function of the temperature ($\alpha_V$) near the initial state
conditions.

\begin{figure}[!h]
\begin{center}
\epsfig{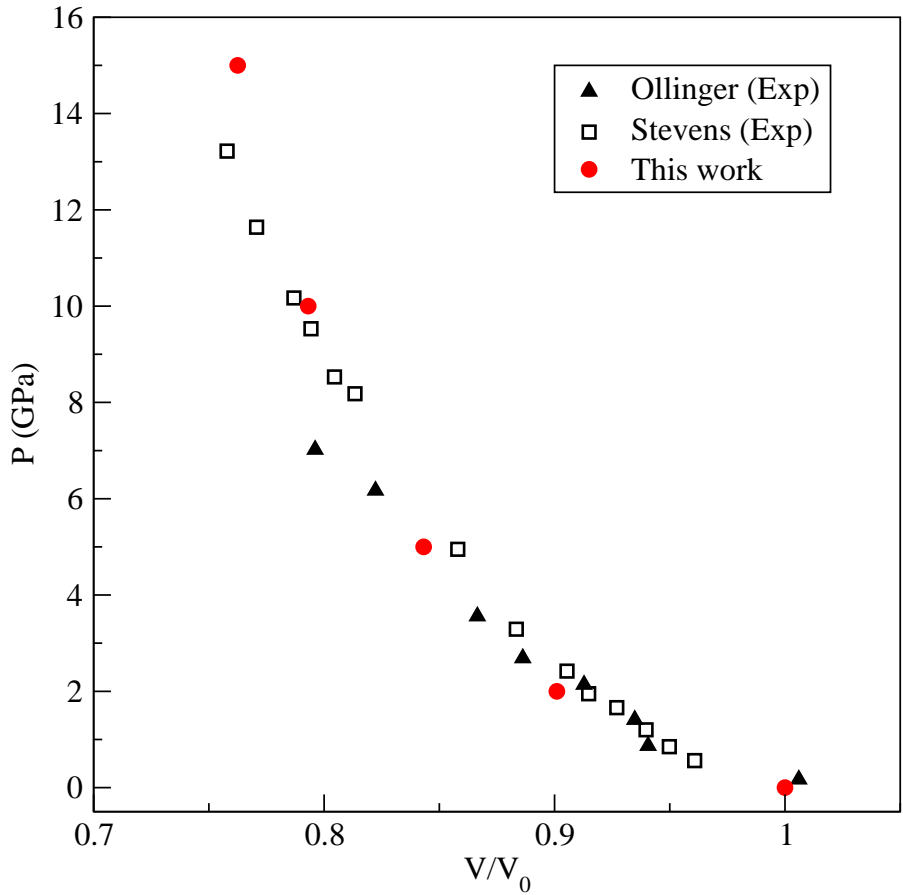}
\end{center}
\caption{TATB isotherm 300~ K: pressure as a function of specific volume. Calculation results are compared to experimental ones from Olinger \cite{olinger76} and Stevens \cite{stevens08}.} \label{FigisoT}
\end{figure}

To analyze the behavior of the potential under pressure, we 
performed 4 NPT simulations at 300~K at different pressures~$P$ (2, 5, 10
and 15 GPa). Results are displayed in Table~\ref{ResisoT}, and
compared to experimental results from Olinger~\cite{olinger76} and
Stevens~\cite{stevens08} in Figure~\ref{FigisoT}. The quoted experimental
results have been obtained by compression of TATB powder, which
roughly corresponds to hydrostatic compression of the
monocrystal. According to Stevens, discrepancies observed between the
results from the two authors come from the hypothesis made to obtain
the cell parameters.
Figure~\ref{FigisoT} shows that numerical results are in good agreement with experimental ones. Rai {\it et al.} have tested their potential up to 7 GPa. We show that this potential gives satisfying qualitative results up to 15 GPa. A careful investigation however reveals that the curvature of the P-V isotherm is not well reproduced, the curve being too convex. This is consistent with the fact that the computed bulk modulus~$K_0$ is smaller than the experimental one.

\begin{table}[h!]
\centering
\begin{tabular}{|c|c|c|}
\hline
P (GPa) & V/V$_0$ & V (cm$^3$.g$^{-1}$) \\
\hline
10$^{-4}$ & 1.0 & 0.517 \\
2.0 & 0.901 & 0.467 \\
5.0 & 0.843 & 0.436 \\
10.0 & 0.793 & 0.410 \\
15.0 & 0.764 & 0.395 \\
\hline
\end{tabular} \caption{Calculated volumes along the isotherm 300 K. Statistical uncertainties are under 0.1 \%.}
\label{ResisoT}
\end{table}

\begin{figure}[!h]
\begin{center}
\epsfig{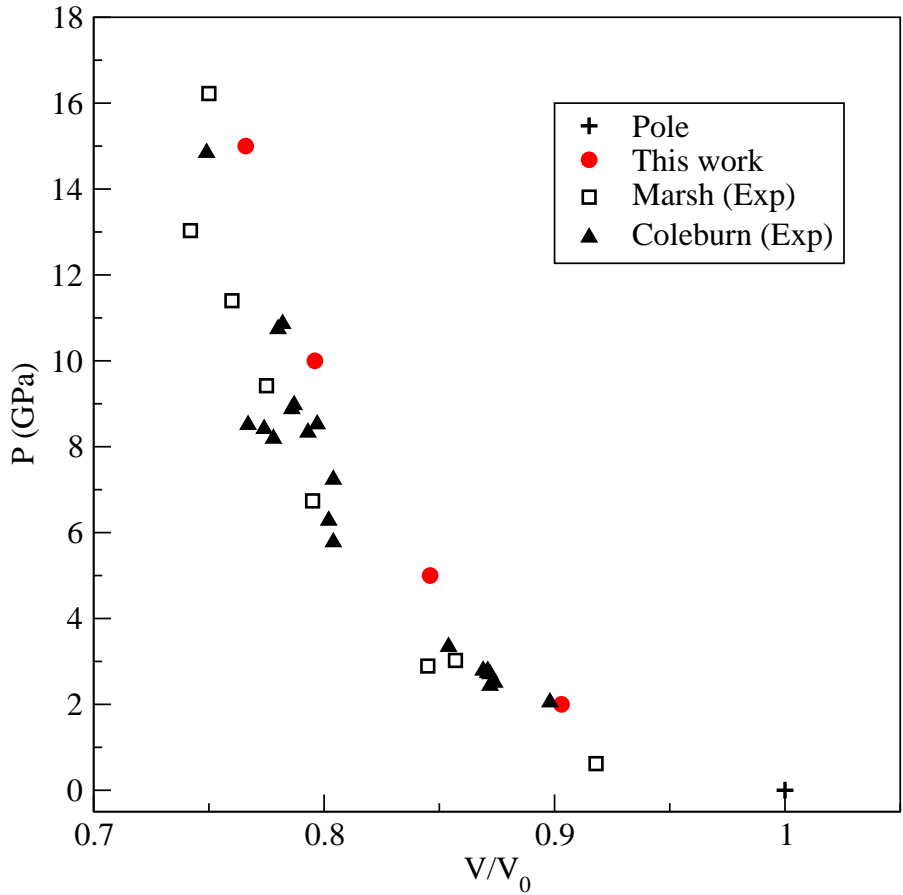}
\end{center}
\caption{Hugoniot of neat TATB. Calculation results are compared to experimental ones from Coleburn \cite{coleburn66} and Marsh \cite{marsh}.} \label{Fighugo}
\end{figure}

To calculate the Hugoniot curve of neat TATB, we performed 4 simulations using the SCA method described in Section~\ref{sec:SCA}, at $P = 2$, 5, 10 and 15~GPa. Results are shown in Table~\ref{Reshugo} and Figure~\ref{Fighugo}, and are compared to experimental results from Coleburn {\it et al.} \cite{coleburn66} and Marsh \cite{marsh}. The latter Hugoniot experimental measurements were obtained from compressed powders slightly less dense (1847 kg.m$^{-3}$ and 1876 kg.m$^{-3}$) than monocrystals (1938 kg.m$^{-3}$). Figure \ref{Fighugo} shows that the pressure along the Hugoniot curve for a given compression is slightly overestimated. The difference seems sufficiently small to be corrected by a modification of the parameters of the potential using some potential optimization, as we previously did for nitromethane~\cite{desbiens07}. Table~\ref{Reshugo} demonstrates the accuracy of the SCA method for obtaining Hugoniot states. The statistical uncertainty on the computed temperatures is really small, and the average value of H$_{\rm g}$ is close to zero, knowing that the total energy of such system is typically always over 1000 J.g$^{-1}$.

\begin{table}[h!]
\centering
\begin{tabular}{|c|c|c|c||c|}
\hline
P (GPa) & V/V$_0$ & V (cm$^3$.g$^{-1}$) & T (K) & H$_{\rm g}$ (J.g$^{-1}$) \\
\hline
10$^{-4}$ & 1.0 & 0.5175 & 300.0 & - \\
2.0 & 0.903 & 0.4675 & 342.9 $\pm$ 0.14 & -0.048 $\pm$ 0.15 \\
5.0 & 0.8458 & 0.4377 & 403.9 $\pm$ 0.11 & 0.068 $\pm$ 0.22 \\
10.0 & 0.7962 & 0.4120 & 520.3 $\pm$ 0.11 & -0.001 $\pm$ 0.25 \\
15.0 & 0.7659 & 0.3963 & 652.9 $\pm$ 0.17 & -0.045 $\pm$ 0.36\\
\hline
\end{tabular} \caption{Pressure, compression, volume and temperature of TATB along the Hugoniot curve. The last column gives the average value of H$_{\rm g}$ calculated during the simulation. Statistical uncertainties on volumes are under 0.1 \%.}
\label{Reshugo}
\end{table}

\subsection{Hugoniot curve of detonation products of TATB}

The detonation product mixture of TATB is mainly composed of 8 molecular species (solid carbon, CO$_2$, H$_2$O, CO, N$_2$, H$_2$, NH$_3$ and CH$_4$). To model the global chemical equilibrium occurring in the system, the 4 following independent elementary chemical equilibriums have been considered simultaneously. Those 4 chemical equations have been determined using the Smith and Missen method \cite{smith79}, explained in~\cite{hervouet08}:
\begin{eqnarray}
2 \, \mathrm{CO} & \rightleftarrows& \mathrm{CO}_2 + \mathrm{C}_{\rm solid}\nonumber\\
2 \, \mathrm{NH}_3 & \rightleftarrows& \mathrm{N}_2 + 3 \, \mathrm{H}_2\nonumber\\
\mathrm{CO} + 2 \, \mathrm{NH}_3 & \rightleftarrows& \mathrm{N}_2 + \mathrm{CH}_4 + 
\mathrm{H}_2\mathrm{O} \nonumber\\
\mathrm{CO}_2 + \mathrm{H}_2 & \rightleftarrows& \mathrm{CO} + \mathrm{H}_2\mathrm{O} \nonumber
\end{eqnarray}

\begin{figure}[!h]
\begin{center}
\epsfig{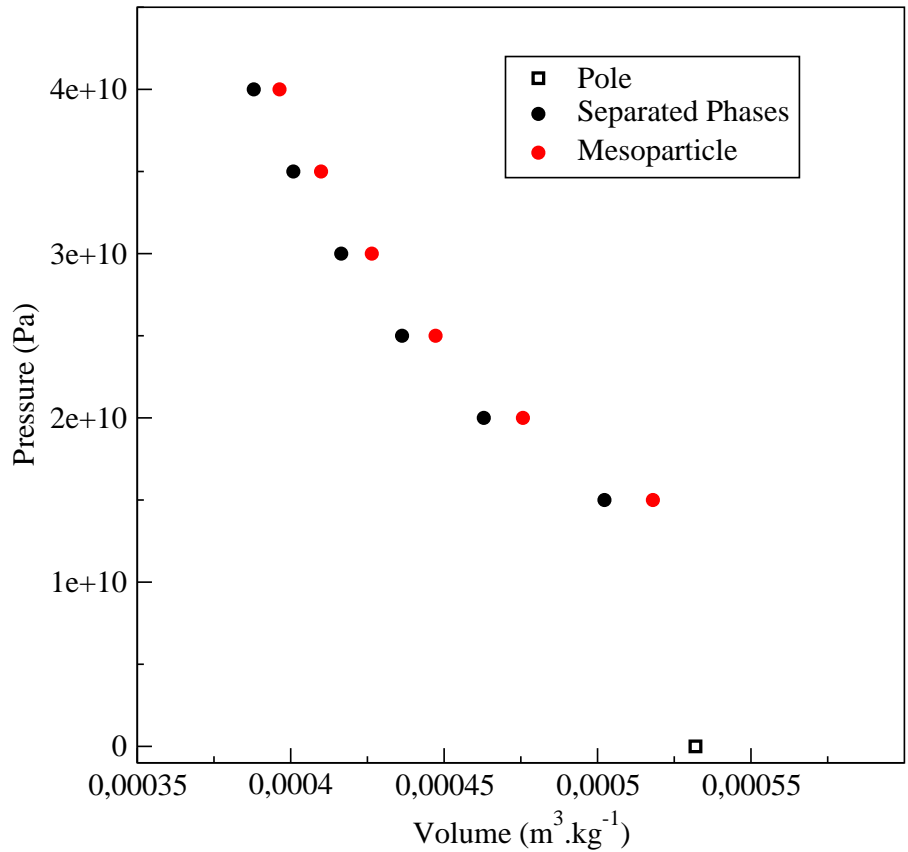}
\end{center}
\caption{Hugoniot of detonation product mixture of TATB in two different cases: either the two phases are considered as completely separated, or the solid phase is modelled through a mesoparticle immersed in the fluid phase.} \label{FigCrussard}
\end{figure}

We computed 6 Hugoniot states of the system at $P$ = 15, 20, 25, 30, 35 and 40~GPa,
using the SCA method in the RxMC ensemble. The initial conditions ($E_0$,$P_0$, $V_0$) are the same as for the unreacted case. These computations were performed under two different hypotheses: either considering the two phases completely separated (using RxMC with the Composite Ensemble~\cite{hervouet08}), or modelling the solid phase as a mesoparticle immersed in the fluid phase (using the RxMC with mesoparticle method~\cite{bourasseau11}). Results are presented in Table~\ref{ResCrussard} and in Figure~\ref{FigCrussard}.

\begin{table}[h!]
\centering
\begin{tabular}{|c||c|c|c|c||c|c|c|c|}
\hline
 & \multicolumn{4}{c||}{Separated phases} & \multicolumn{4}{c|}{Mesoparticle} \\
\hline
P (GPa) & V/V$_0$ & V (cm$^3$.g$^{-1}$) & T (K) & H$_{\rm g}$ (J.g$^{-1}$) & V/V$_0$ & V (cm$^3$.g$^{-1}$) & T (K) & H$_{\rm g}$ (J.g$^{-1}$) \\
\hline
15.0 & 0.9443 & 0.5023 & 2821.5 $\pm$ 0.2 & 0.358 $\pm$ 3.5 & 0.9739 & 0.5180 & 2596.4 $\pm$ 4.9 & -0.035 $\pm$ 7.9 \\
20.0 & 0.8703 & 0.4629 & 2902.8 $\pm$ 0.4 & -0.350 $\pm$ 3.8 & 0.8942 & 0.4757 & 2657.0 $\pm$ 1.4 & -0.517 $\pm$ 5.7 \\
25.0 & 0.8202 & 0.4363 & 2992.7 $\pm$ 0.4 & 0.775 $\pm$ 4.0 & 0.8408 & 0.4472 & 2720.1 $\pm$ 3.2 & -0.268 $\pm$ 3.6 \\
30.0 & 0.7829 & 0.4164 & 3093.4 $\pm$ 1.9 & -0.893 $\pm$ 4.9 & 0.8017 & 0.4264 & 2785.5 $\pm$ 5.9 & 0.997 $\pm$ 5.3 \\
35.0 & 0.7536 & 0.4008 & 3208.5 $\pm$ 1.4 & -0.857 $\pm$ 4.7 & 0.7706 & 0.4099 & 2867.7 $\pm$ 5.6 & 0.199 $\pm$ 4.4 \\
40.0 & 0.7294 & 0.3880 & 3339.1 $\pm$ 4.3 & -0.771 $\pm$ 4.8 & 0.7466 & 0.3971 & 2936.0 $\pm$ 3.9 & 0.399 $\pm$ 3.8 \\
\hline
\end{tabular} \caption{Pressure, compression, volume and temperature of detonation product mixture of TATB along the Hugoniot curve in two different cases: either the two phases are considered as completely separated, or the solid phase is modelled through a mesoparticle immersed in the fluid phase. The last column gives the average value of H$_{\rm g}$ (see equation \ref{Hg}) calculated during the simulation. Statistical uncertainties on volumes are under 0.1 \%.}
\label{ResCrussard}
\end{table}

Table \ref{ResCrussard} shows that the SCA method gives also good results in this case: The values of H$_{\rm g}$ are really close to zero. The uncertainties on the temperatures are around a few Kelvins, and uncertainties on H$_{\rm g}$ are around a few J.g$^{-1}$, what appears satisfying knowing that the energy of such systems is typically superior to several hundreds of J.g$^{-1}$. This is clearly satisfying, even if uncertainties are higher than for neat TATB. This is due to the fact that there are intrinsically more fluctuations on the detonation product mixture than for neat TATB since the system is in a fluid phase, and its chemical composition varies.

\begin{figure}[!h]
\begin{center}
\epsfig{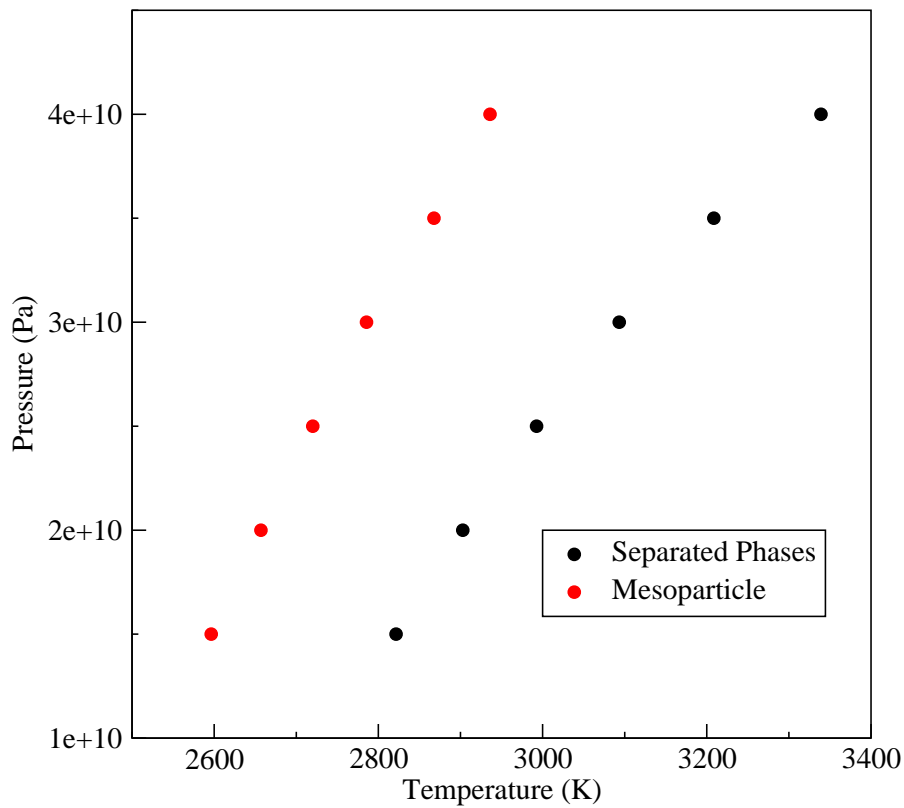}
\end{center}
\caption{Pressure/Temperature evolution along the Hugoniot curve of detonation products of TATB in two different cases: either the two phases are considered as completely separated, or the solid phase is modelled through a mesoparticle immersed in the fluid phase.} \label{FigCrussard2}
\end{figure}

\begin{figure}[!h]
\begin{center}
\epsfig{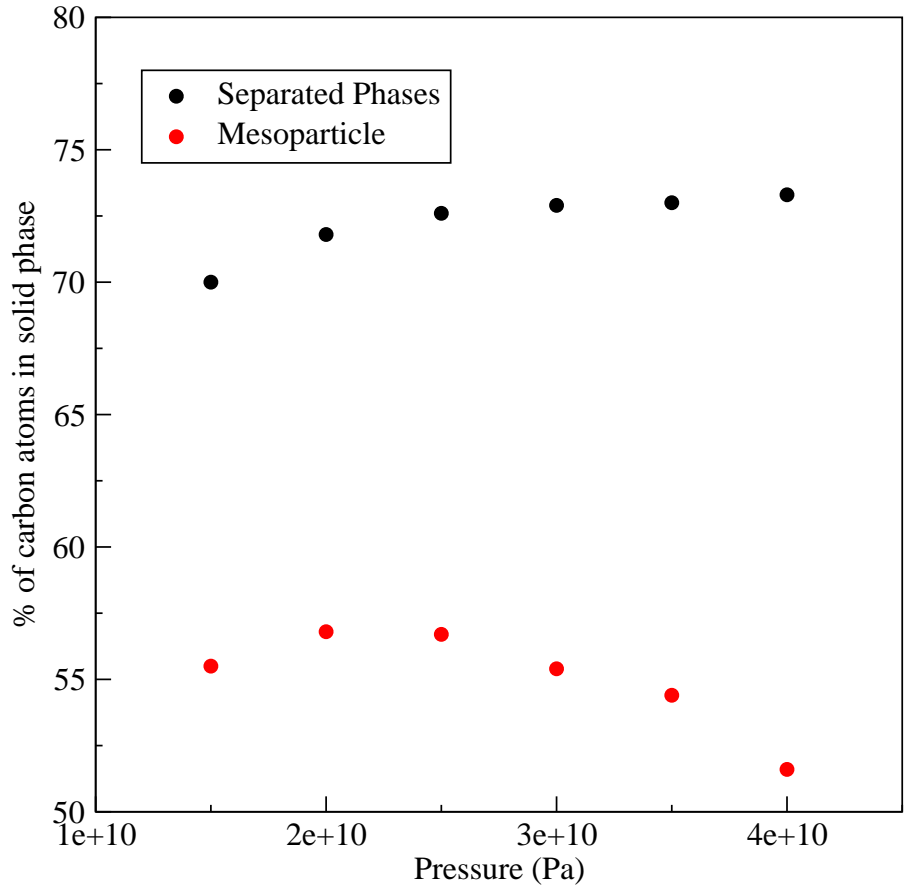}
\end{center}
\caption{Evolution of the percentage of carbon atoms in the solid phase along the Hugoniot of detonation product mixture of TATB in two different cases: either the two phases are considered as completely separated, or the solid phase is modelled through a mesoparticle immersed in the fluid phase.} \label{FigCrussard3}
\end{figure}

Figure \ref{FigCrussard} shows that significant differences (up to 4.3 \% on the calculated volume at a given pressure P) appear between the results obtained with the two different assumptions on the modelling of the solid phase (separated phases \textit{vs.} mesoparticle immersed in the fluid phase). This has already been shown and explained in a previous paper \cite{bourasseau11}. The discrepancies can be attributed to the expansion of the solid phase, which needs more energy in the second case because mesoparticles have to overcome the fluid pressure to grow. As a consequence, the chemical equilibrium of the system is displaced towards less carbon atoms in the solid phase for a given pressure and temperature. The consequence of this difference on the Hugoniot curve is difficult to explain, but Figures~\ref{FigCrussard2} and~\ref{FigCrussard3} show that the evolution of the temperature and the amount of carbon in the solid phase along the Hugoniot are really different in the two situations. The differences in the computed temperatures with the two hypotheses can reach 12 \% at high pressures. The difference in the amount of carbon atoms in the solid phase can reach 30 \% at 40 GPa, and we also observe that the qualitative evolution of the amount of solid carbon is totally different. This is a supplementary evidence that the heterogeneity of the system must be taken into account.
 
%------------- conclusion -----------------
\section{Conclusions}
\label{sec:conclusion}

This work has shown that the "Sampling Constraints in Average" method can be used to calculate, in the same framework, the Hugoniot curves of neat TATB and of detonation products of TATB, with efficiency and reliability. The statistical uncertainties obtained both on converged temperatures and average values of $H_{\rm g}$ are more than reasonable. Besides, the parameters of the method are easy to determine. Moreover, derivative thermodynamic properties can be computed accurately.

The potential we used to model the neat TATB, proposed by Rai \etal~\cite{rai08}, allows to reproduce correctly the direct thermodynamic properties of TATB at the pole conditions. Nevertheless it fails to reproduce quantitatively the derivative properties, and in particular the compressibility of the system. As a consequence, the 300~K isotherm and the Hugoniot curve are both too convex. This could probably be corrected with an appropriate modification of the parameters of the potential through some optimization procedure \cite{desbiens07}. This work is currently in progress.

We used the specific RxMC method \cite{bourasseau11} to obtain the chemical equilibrium of the detonation product mixture of TATB, including solid carbon clusters. In this method, the all-atoms carbon clusters are replaced by mesoparticles. Combining this version of the RxMC method and the SCA method allowed us to compute the Hugoniot curve of the detonation product mixture of TATB. As expected, the results differ significantly from the results obtained with the composite ensemble, where the two phases are considered as completely separated. This is an evidence of the fact that the heterogeneity of the system (\textit{i.e.} the fact that the carbon clusters are immersed in the detonation products fluid) must be taken into account. This is particularly important to calculate the amount of carbon atom included in the solid phase.

To conclude, let us emphasize that the detonation velocities computed from the results of this work cannot be compared with experimental detonation velocities because the equation of state used to model the solid phase is not representative of a cluster phase. Nevertheless, we are currently working on the parametrization of a carbon cluster equation of state based on molecular dynamics simulation results obtained with the LCBOPII potential \cite{chevrot09}. Using this new equation of state, we anticipate a more accurate prediction of detonation velocity of TATB. 

\section{Acknowledgements}

All Monte Carlo simulations have been performed with the Gibbs code from IFP, CNRS and the Université Paris-Sud \cite{ungerer}. 
G.S. acknowledges the support of the French Ministry of 
Education through the grant ANR-09-BLAN-0216-01 (MEGAS).

\bibliographystyle{unsrt}
\bibliography{biblio}

\end{document}